\documentclass[aps,pra,showpacs,twocolumn,floatfix]{revtex4}
\usepackage{amssymb}

\usepackage{graphicx}
\usepackage{bm}

\begin{document}

\title{Non-linear amplification of small spin precession using long range dipolar interactions}
\author{M. P. Ledbetter, I. M. Savukov, and M. V. Romalis}
\affiliation{Department of Physics, Princeton University,
Princeton, New Jersey 08544}
\date{\today}

\begin{abstract}

In measurements of small signals using spin precession the
precession angle usually grows linearly in time. We show that a
dynamic instability caused by spin interactions can lead to an
exponentially growing spin precession angle, amplifying small
signals and raising them above the noise level of a detection
system.   We demonstrate amplification by a factor of greater than
8 of a spin precession signal due to a small magnetic field
gradient in a spherical cell filled with hyperpolarized liquid
$^{129}$Xe. This technique can improve the sensitivity in many
measurements that are limited by the noise of the detection
system, rather then the fundamental spin-projection noise.

\end{abstract}

\maketitle

Observation of spin precession signals forms the basis of such
prevalent experimental techniques as NMR and EPR.  It is also used
in searches for physics beyond the Standard
Model~\cite{Regan,Bear,RomalisEDM,Hunter96} and sensitive
magnetometery~\cite{Kominis}. Hence, there is significant interest
in the development of general techniques for increasing the
sensitivity of spin precession measurements. Several methods for
reducing spin-projection noise using quantum non-demolition
measurements have been explored \cite{Bigelow,Mabuchi} and it has
been shown that in some cases they can lead to improvements in
sensitivity \cite{Budker,Lukin}. In this Letter we demonstrate a
different technique that increases the sensitivity by amplifying
the spin precession signal rather than reducing the noise.

The amplification technique is based on the exponential growth of
the spin precession angle in systems with a dynamic instability
caused by collective spin interactions. Such instabilities can be
caused by a variety of interactions, for example, magnetic dipolar
fields in a nuclear-spin-polarized liquid
\cite{Warren,Jeener99,Sauer} or electron-spin polarized gas
\cite{Vasilyev}, spin-exchange collisions in an alkali-metal vapor
\cite{Klipstein} or mixtures of alkali-metal and noble-gas atoms
\cite{Kornack}. This amplification technique can be used in a
search for a permanent electric dipole moment in liquid $^{129}$Xe
\cite{Romalis2001}. It is also likely to find applications in a
variety of other systems with strong dipolar interactions, such as
cold atomic gases \cite{Pfau} and polar molecules \cite{Demille}.

Consider first an ensemble of non-interacting spins with a
gyromagnetic ratio $\gamma$ initially polarized in the $\hat{x}$
direction and precessing in a small magnetic field $B_z$. The spin
precession signal $\langle S_y \rangle = \gamma \langle S_x
\rangle B_z t$ grows linearly in time for $\gamma  B_z t \ll 1$.
The measurement time $t_m$ is usually limited by spin relaxation
processes and determines, together with the precision of  spin
measurements $\delta (\langle S_y \rangle)$, the sensitivity to
the magnetic field $B_z$
\begin{equation}
\delta B_z = \frac{\delta (\langle S_y \rangle)}{\gamma \langle
S_x \rangle t_m}
\end{equation}
or any other interaction coupling to the spins. In the presence of
a dynamic instability, the initial spin precession away from a
point of unstable equilibrium can be generally written as $\langle
S_y \rangle = \gamma \langle S_x \rangle  B_z \sinh(\beta
t)/\beta$, where $\beta$ is a growth rate characterizing the
strength of spin interactions. The measurement uncertainty is now
given by
\begin{equation}
\delta B_z = \frac{\delta (\langle S_y \rangle) \beta}{\gamma
\langle S_x \rangle \sinh(\beta t_m)}.
\end{equation}
Hence, for the same uncertainty in the measurement of $ \langle
S_y \rangle$, the sensitivity to $B_z$ is improved by a factor of
$G = \sinh(\beta t_m)/\beta t_m$. It will be shown that quantum
(as well as non-quantum) fluctuations of $\langle S_y \rangle$ are
also amplified, so this technique cannot be used to increase the
sensitivity in measurements limited by the spin-projection noise.
However, the majority of experiments are not limited by quantum
fluctuations. For a small number of spins the detector sensitivity
is usually insufficient to measure the spin-projection noise of
$N^{1/2}$ spins, while for a large number of particles the dynamic
range of the measurement system is often insufficient to measure a
signal with a fractional uncertainty of $N^{-1/2}$. Amplifying the
spin-precession signal before detection reduces the requirements
for both the sensitivity and the dynamic range of the measurement
system. Optical methods allow efficient detection of electron
spins  and some nuclear spins \cite{RomalisEDM} in atoms or
molecules with  convenient optical transitions. However, for the
majority of nuclei optical detection methods are not practical and
magnetic detection, using, for example, magnetic resonance force
microscopy, has not yet reached the sensitivity where it is
limited by the spin projection noise \cite{Sidles,Thurber}.
Therefore, non-linear amplification can lead to particularly large
improvements in precision measurements relying on nuclear spin
precession.

Here we use long-range magnetic dipolar interactions between
nuclear spins that lead to exponential amplification of spin
precession due to a magnetic field gradient
\cite{Jeener99,Romalis2001,Ledbetter2002}. It has also been shown
that long-range dipolar fields in conjunction with radiation
damping due to coupling with an NMR coil lead to an increased
sensitivity to initial conditions and chaos \cite{Lin}. To amplify
a small spin precession  signal above detector noise it is
important that the dynamic instability involves only spin
interactions, since instabilities caused by the feedback from the
detection system would couple the detector noise, such as the
Johnson noise of the NMR coil, back to the spins. We measure spin
precession using SQUID magnetometers that do not have a
significant back-reaction on the spins and show that under well
controlled experimental conditions the dynamic instability due to
collective spin interactions can be used to amplify small spin
precession signals in a predictable way.

Our measurements are performed in a spherical cell containing
hyperpolarized liquid $^{129}$Xe (Fig.~1). Liquid $^{129}$Xe has a
remarkably long spin relaxation time \cite{Romalis2001} and the
spin dynamics is dominated by the effects of long-range magnetic
dipolar fields. In the spherical geometry an analytic solution can
be found using a perturbation expansion in a nearly uniform
magnetic field $H_{0}$ \cite{Romalis2001,Ledbetter2004}. We are
primarily interested in the first-order longitudinal magnetic
field gradient $g$, $\mathbf{H}=(H_{0}+gz)\mathbf{\hat{z}}$, but
will also consider other magnetic field gradients which inevitably
arise due to experimental imperfections. For longitudinal
gradients that preserve cylindrical symmetry the magnetization
profile can be expanded in a Taylor series,
\begin{equation}
\mathbf{M}(\mathbf{r},t)= \mathbf{M}_{0}+M_{0}\sum_{i,k}
\mathbf{m}^{(i,k)}(t)\frac{z^i (x^2+y^2)^k}{R^{i+2k}},
\label{linexpan}
\end{equation}
where $R$ is the radius of the cell. Only gradients of the
magnetization create dipolar fields in a spherical cell, for
example, a linear magnetization gradient $\mathbf{m}^{(1,0)}$
creates only a linear dipolar magnetic field, which, in the
rotating frame, is given by
\begin{equation}
\mathbf{B}_d^{(1,0)}=\frac{8\pi M_0 z}{15R}\left\{
m_{x}^{(1,0)},m_{y}^{(1,0)},-2 m_{z}^{(1,0)}\right\}.
\label{fieldgrad}
\end{equation}
The time evolution of the magnetization is determined by the Bloch
equations $d\mathbf{M}/dt=\gamma \mathbf{M}\times (
\mathbf{B}_d+\mathbf{H})$. If the magnetization is nearly uniform,
$\mathbf{m}^{(i,k)} \ll 1$, they can be reduced to a system of
linear first-order differential equations for
$\mathbf{m}^{(i,k)}$.

\begin{figure}[tbp]
\includegraphics[width=3.0in]{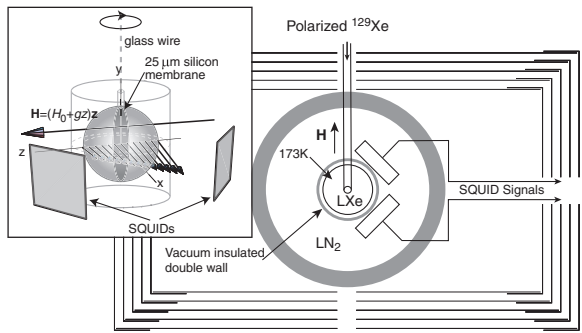}
\caption{Low field NMR setup (view from above). Polarized liquid
$^{129}$Xe is contained in a spherical cell maintained at 173K by
flowing N$_{2}$ gas through a vacuum insulated column.
High-$T_{c}$ SQUIDs are submerged in LN$_{2}$ contained in a glass
dewar.  Inset: configuration of the SQUIDs, applied magnetic
field, the magnetization, and the rotatable membrane.}
\label{Setup}
\end{figure}

We consider first the simplest case when only the linear field
gradient $g$ is present and the initial uniform magnetization
$M_{0}$  is tipped into the $\hat{x}$ direction of the rotating
frame by a $\pi/2$  pulse. Substituting Eqns.~(\ref{linexpan}) and
(\ref{fieldgrad}) into the Bloch equations we find that only
linear magnetization gradients grow as long as $\mathbf{m}^{(i,k)}
\ll 1$, in particular, $m_y^{(1,0)}$ is given by
\begin{eqnarray}
&&m_{y}^{(1,0)}(t)=-\frac{\gamma g R}{\beta } \sinh (\beta t),
\label{grad} \\
&&\beta =\frac{8\sqrt{2}\pi }{15}M_{0}\gamma. \label{beta}
\end{eqnarray}
Here $\beta $ is proportional to the strength of the long-range
dipolar interactions.  We measure $m_{y}^{(1,0)}$  experimentally
by placing two SQUID detectors near the spherical cell as
illustrated in Fig.~\ref{Setup} and measuring the phase difference
$\Delta \phi$ between the NMR signals induced in the two SQUIDs.
For small $m_{y}^{(1,0)}$, $\Delta \phi =\zeta m_{y}^{(1,0)}$,
where $\zeta $ is a numerical factor that depends on the geometry,
for our dimensions $\zeta =0.46\pm 0.01$. Thus, the phase
difference $\Delta \phi$ is proportional to the applied magnetic
field gradient $g$ and grows exponentially in time, increasing the
sensitivity to $g$ by a factor $G=\sinh(\beta t)/\beta t$. For
$M_{0} = 100$ $\mu$G, which is easy to realize experimentally with
hyperpolarized $^{129}$Xe, $\beta = 1.75~{\rm sec}^{-1} $, so that
a very large amplification factor can be achieved in a short time,
for example $G = 360$ after 5 seconds.

\begin{figure}
  \includegraphics[width=3.0in]{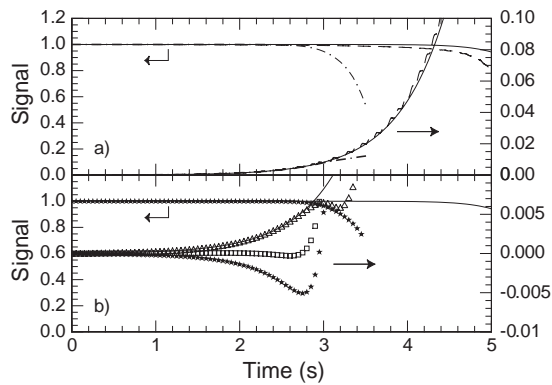}\\
  \caption{Numerical simulations \cite{Ledbetter2004} of the SQUID
  signal (left axis) and the phase difference between SQUIDs (right
  axis) for $M_0 = 100\thinspace\mu$G and a small longitudinal field
   gradient $g = 0.1\thinspace\mu$G/cm (solid lines).
  a)  An additional larger transverse gradient $g_\perp =
2\thinspace\mu$G/cm (dashed line) or a second-order longitudinal
gradient $g_2 =1\thinspace \mu$G/cm$^2$ (dash-dot) do not affect
the phase difference until the SQUID signal begins to  decay.
 b) Effects of an additional $z$-odd third-order longitudinal gradient  $g_3 = 0.8 \thinspace \mu$G/cm$^3$ (squares). Stars show the
  phase evolution in the presence of $g_3$ but for $g=0$. The difference between the phase
  for $g = 0.1\thinspace\mu$G/cm and $g=0$ (triangles) follows the solid line corresponding
  to the pure linear gradient $g$ until the magnetization begins to collapse. The third-order gradient
  generates a background phase that can be subtracted to determine
  a  change in $g$ between successive measurements.
  }\label{simfig}
\end{figure}

One of the main challenges to realizing such high gains is to
achieve sufficient control over the initial conditions and
non-linear evolution of the system, so that the dynamic
instability gives rise to a phase difference $\Delta \phi$ that
remains proportional to $g$ even in the presence of various
experimental imperfections. We developed a set of numerical and
analytical methods for analyzing these effects
\cite{Ledbetter2004}. Since our goal is to achieve very high
sensitivity to a small first-order longitudinal magnetic field
gradient $g$, we generally assume that it is smaller than other
gradients that are not measured directly. We find that  the
presence of transverse or higher order longitudinal gradients as
well as initial magnetization inhomogeneities cause an abrupt
non-linear decay of the overall magnetization. The time until the
decay $t_c$  depends on the size of the inhomogeneities relative
to $M_0$ and limits the achievable gain to $\sinh(\beta t_c)/\beta
t_c$. Inhomogeneities of the applied field symmetric with respect
to the $z$ direction do not change the evolution of $\Delta \phi$,
which remains proportional to $g$ until the collapse of the
magnetization, as shown in Fig.~\ref{simfig}a. Higher order
$z$-odd longitudinal gradients do generate a phase difference
(Fig.~\ref{simfig}b). However, the contributions of different
magnetic field gradients to the phase difference add linearly as
long as $\mathbf{m}^{(i,k)} \ll 1$ and the effects of higher order
odd gradients can be subtracted if they remain constant, as
illustrated in Fig.~\ref{simfig}b. While higher order
magnetization gradients can grow with a time constant up to 2.5
times faster than the first-order gradient, it can be shown using
a perturbation expansion that the first moment of the
magnetization $d=\int zM_{y} dV$ always grows with an exponential
constant given by Eq.~(\ref{beta}) and is proportional to the
first moment of the magnetic field $b=\int zB_{z} dV$.  The phase
difference between the SQUID signals is approximately proportional
to the first moment of the magnetization $d$  and is not
significantly affected by the growth of higher order gradients.
For example, in Fig.~\ref{simfig}b) the overall signal decays at
about 3 sec due to large first and third-order magnetization
gradients but the phase difference $\Delta \phi$ remains much less
than 1.

Hence, the phase difference $\Delta \phi$ can be used to measure a
very small linear gradient $g$ in the presence of larger
inhomogeneities if all magnetic field and magnetization
inhomogeneities are much smaller than $M_0$. The ultimate
sensitivity is limited by fluctuations of the gradients between
successive measurements. In addition to fluctuations of $g$, which
is the quantity being measured, the phase difference will be
affected by the fluctuations in the  initial magnetization
gradients $m_y^{(1,0)}$ and $m_z^{(1,0)}$ and, to a smaller
degree, higher order $z$-odd gradients of the magnetic field and
the magnetization. In particular, fluctuations of $m_y^{(1,0)}$
and $m_z^{(1,0)}$, either due to spin-projection noise or
experimental imperfections, set a limit on the magnetic field
gradient sensitivity on the order of $\delta g = 8 \pi\sqrt{2} M_0
\delta m_y^{(1,0)}/15R$ and similar for $\delta m_z^{(1,0)}$. The
shot noise fluctuations of $^{129}$Xe magnetization generate a
magnetic field gradient on the order of $10^{-13}$~G/cm.

Hyperpolarized $^{129}$Xe is produced using the standard method of
spin exchange optical pumping \cite{Romalis2001,Bastian}. The
polarized gas is condensed in a spherical glass cell held at 173~K
as shown in Fig.~\ref {Setup}. The cell, with an inner radius
$R=0.55$ cm, is constructed from two concave hemispherical lenses
glued together with UV curing cement. Inside the cell is an
octagonal silicon membrane 25 $\mu$m thick, with a diameter of
1.05 cm. The membrane is connected to a stepper motor outside the
magnetic shields via a ~0.2 mm glass wire to mix the sample,
ensuring uniformity of the polarization. In addition to mixing the
sample, the membrane inhibits convection across the cell due to
small temperature gradients which can wash out the longitudinal
gradient of the magnetization. A set of coils inside the shields
create a 10 mG uniform magnetic field and allow application of RF
pulses and control of linear and quadratic magnetic field
gradients. The NMR signal is detected using high-$T_{c}$ SQUID
detectors. The pick-up coil of each SQUID detector is an 8$ \times
8$ mm square loop located approximately 1.6 cm from the center of
the cell and tilted by $\pm 45^{\circ }$ relative to the magnetic
field.

\begin{figure}
  \includegraphics[width=2.8in]{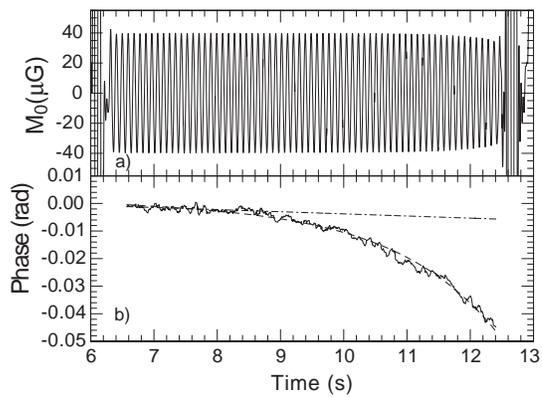}\\
  \caption{a) Oscillating transverse magnetization following a $\pi$/2 pulse.  After the signal
  drops to 90\% of its initial value a second pulse is applied to
  realign the magnetization with the longitudinal direction. b)
  Phase difference between the SQUID signals. Overlaying the data (dashed line) is a
  fit based on Eq. (\ref{grad}).  The dash-dot line is the expected phase evolution in the
  non-interacting case, illustrating that the signal would barely be detectable.}\label{bigpulse}
\end{figure}

In our experimental system, the time scale of the dipolar
interactions is much smaller than the spin relaxation time or the
time needed to polarize a fresh sample of $^{129}$Xe.  In order to
make multiple measurements on a single sample of polarized xenon,
we first apply a $\pi/2$ pulse and monitor in real time the SQUID
signals. When the NMR signal drops to 90\% of its initial value, a
second $\pi/2$ pulse is applied, realigning the magnetization with
the holding field. The silicon membrane is then oscillated back
and forth to erase the magnetization inhomogeneities developed in
the previous trial.

Fig.~\ref{bigpulse}a) shows the oscillating transverse
magnetization and Fig.~\ref{bigpulse}b) shows the phase difference
between the two SQUID signals. We determine the value of $\beta$
from the magnitude of the NMR signal and fit the phase difference
to Eq. (\ref{grad}) with $g$ as the only free parameter. The
dash-dot line shows the expected evolution of the phase difference
for the same gradient in the absence of dipolar interactions,
demonstrating that without amplification the phase difference
would be barely above the noise level of the detection system. For
this measurement the phase is amplified by a factor of 9.5 before
the magnetization drops to 90\% of its initial value.

\begin{figure}[t]
  \includegraphics[width=2.8in]{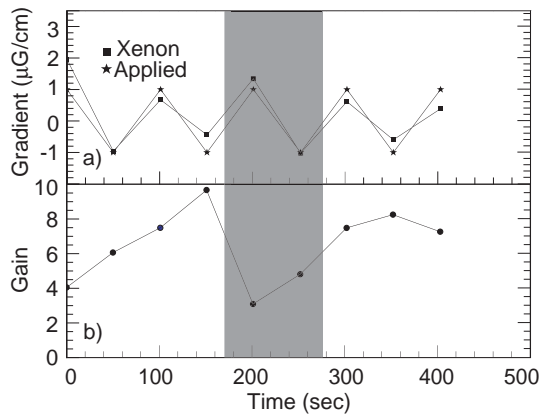}\\
  \caption{a) Measurement of a small gradient $g$ alternated between successive trials.
  Stars show the applied linear gradient, squares show
  the gradient measured using non-linear spin evolution.  b)  Gain $G$ associated with non-linear spin
  evolution. The gain drops when the sample is not mixed in the shaded region, demonstrating
  the significance of initial magnetization inhomogeneities.}\label{gradosc}
\end{figure}

By applying a series of double $\pi/2$ pulses we can make repeated
measurements of the magnetic field gradient.  Fig.~\ref{gradosc}a)
shows data where the applied longitudinal gradient is oscillated
with an amplitude of 1 $\mu$G/cm between trials. The stars show
the applied gradient, the squares show the gradient measured by
the non-linear spin evolution, indicating that the amplified
signal follows the applied gradient. Slight differences between
the two curves are due to noise in the magnetic field gradient as
well as possible imperfections in the erasing of magnetization
gradients between successive trials. Fig.~\ref{gradosc}b) shows
the gain parameter for the same data set. We associate the rising
gain at the beginning of the data set with a decay of the
magnetization inhomogeneities developed during collection of
liquid $^{129}$Xe in the cell. In the shaded region of the plot we
did not mix the magnetization with the membrane before the
measurement, resulting in a drop of the gain as well. Numerical
simulations indicate that the gain is likely limited by higher
order gradients, for example a second-order magnetic field
gradient on the order of 1~$\mu$G/cm$^2$, which can not be
excluded based on our mapping of ambient fields, is sufficient to
limit the gain to about 10.

In conclusion, we have demonstrated that non-linear dynamics
arising from long range dipolar interactions can be used to
amplify small spin precession signals, improving the
signal-to-noise ratio under conditions where limitations of the
spin detection system dominate the spin projection noise. By
amplifying the signal before detection, this technique reduces the
requirements on the sensitivity of the detection technique as well
as its dynamic range. In addition to precision measurements, this
technique can potentially be used to amplify small spin precession
signals in various MRI applications, allowing, for example, direct
detection and imaging of the magnetic fields generated by neurons
with MRI \cite{Xiong}. Initial inhomogeneities of the
magnetization, caused, for example, by very slight differences of
$T_1$ in tissues, can also be amplified. We thank DOE, NSF, the
Packard Foundation and Princeton University for support of this
project.

\end{document}